\documentstyle[psfig,preprint,aps]{revtex}

\tightenlines

\def\beq{\begin{equation}}
\def\eeq{\end{equation}}
\def\bea{\begin{eqnarray}}
\def\eea{\end{eqnarray}}

\begin{document}
\hoffset-1cm
\draft

\title{Non-Perturbative Dilepton Production from a Quark-Gluon Plasma 
\footnote{Supported by BMBF, GSI Darmstadt, DFG, and Humboldt foundation}}

\author{Munshi G. Mustafa$^1$\footnote{Humboldt fellow and on leave from Saha 
Institute of Nuclear Physics, 1/AF Bidhan Nagar, Calcutta 700 064, India}, 
Andreas Sch\"afer$^2$, and Markus H. Thoma$^1$\footnote{Heisenberg fellow}}
\address{$^1$Institut f\"ur Theoretische Physik, Universit\"at Giessen, 35392 
Giessen, Germany}
\address{$^2$Institut f\"ur Theoretische Physik, Universit\"at Regensburg, 
93040 Regensburg} 

\date{\today}

\maketitle

\begin{abstract}

The dilepton production rate from the quark-gluon plasma is calculated 
from the imaginary part of the photon self energy using a quark propagator 
that contains the gluon condensate. The low mass dilepton rate obtained in 
this way exhibits interesting structures (peaks and gaps), which might be
observable at RHIC and LHC.

\end{abstract} 

\bigskip

\medskip


\narrowtext
\newpage

\section{Introduction}

Thermal dileptons emitted from the fireball in ultrarelativistic heavy ion 
collisions might serve as a promising signature for the quark-gluon (QGP)
formation in such collisions \cite{ref1}. In contrast to hadronic signals 
dileptons and photons carry direct information about the early phase of
the fireball, since they do not interact with the surrounding medium after 
their production \cite{ref2}. Therefore they can be used as a direct probe
for the QGP. Unfortunately there is a huge background coming from hadronic
decays. Hence it would be desirable to have some specific features in the 
dilepton spectrum which could signal the presence of deconfined matter.
Indeed perturbative calculations \cite{ref3,ref4} have shown distinct
structures (van Hove singularities, gaps)
in the production rate of low mass dileptons caused by non-trivial
in-medium quark 
dispersion relations. Unfortunately such calculations are not reliable at
temperatures within reach of heavy ion collisions, where the coupling
constant is not small. Moreover, perturbative calculations of the dilepton
rate seem not to converge even in the small coupling limit \cite{ref5}.

Lattice QCD, on the other hand, is not capable so far to compute
dynamical quantities such as the dilepton production rate. However,
lattice calculations provide clear evidence for the existence of 
non-perturbative
effects, e.g. effective parton masses \cite{ref6}, hadronic correlators 
\cite{ref7}, and the gluon condensate \cite{boyd}, above the phase transition. 
Recently QCD Green functions at finite temperature, 
which take the presence of a gluon condensate in the QGP into account
\cite{markus,ref10}, have been constructed. 

In the present paper we will study the influence of this non-perturbative
effect on the dilepton production rate. For this purpose, we will calculate
the dilepton rate from the imaginary part of the photon self energy using an
effective quark propagator that contain the gluon condensate. We will find
similar structures as in the perturbative case coming from the quark dispersion 
relation which follows from the pole of the effective quark propagator. These 
structures might serve as an unique signature for the presence of deconfined,
collective quarks in ultrarelativistic heavy ion collisions.

In the next section we will shortly review the basic ideas and results for the
thermal quark propagator in the presence of a gluon condensate. In section 3
we will present our calculations of the dilepton production rate, before we
will discuss our results in section 4. 

\section{Quark propagation and gluon condensate}

The effective quark propagator 
follows from the quark self energy containing the gluon condensate
\cite{markus,lavelle}. The following gauge independent result for this 
propagator at finite temperature has been found: 
\begin{equation}
\tilde S(L) = \frac{\gamma_0-{\hat l}\cdot \vec \gamma}{2D_{+}(L)} +
\frac{\gamma_0+{\hat l}\cdot \vec \gamma}{2D_{-}(L)} \ , \label{effprop}
\end{equation}
where
\begin{equation}
D_\pm(L) = (-l_0 \pm l)(1+a) - b , \label{cond}
\end{equation}
and $L$ is the fermionic four-momentum defined as 
$L= (l_0,\vec l\>)$. The expressions for $a$ and $b$ are obtained
in terms of the chromoelectric and chromomagnetic condensates as~\cite{markus}
\begin{eqnarray}
a &=& - \frac{g^2}{6} \frac {1}{L^6} \left [ \left ( \frac {1}{3} l^2
- \frac{5}{3}l_0^2\right )\langle {\cal E}^2 \rangle_T - \left ( \frac{1}{5}l^2
- l_0^2 \right ) \langle {\cal B}^2 \rangle_T \right ] , \nonumber \\
b &=& - \frac{4}{9}g^2 \frac {l_0}{L^6} \left [ l_0^2\langle {\cal E}^2 
\rangle_T 
+ \frac{1}{5}l^2\langle {\cal B}^2 \rangle_T \right ] ,  \label{ab}
\end{eqnarray}
where $g^2=4\pi \alpha_s$.
The in-medium chromoelectric, $\langle {\cal E}^2\rangle_T$, and 
chromomagnetic condensates, $\langle {\cal B}^2\rangle_T$, come from
the non-perturbative
longitudinal and transverse gluon propagators entering the
quark self energy~\cite{markus}. In Minkowski space these
condensates can be expressed in terms of the space like ($\Delta_\sigma$) 
and time
like ($\Delta_\tau$) plaquette expectation values measured on a 
lattice~\cite{boyd} by~\cite{markus}
\begin{eqnarray}
\frac{\alpha_s}{\pi} \langle {\cal E}^2\rangle_T &=& \frac{4}{11} T^4 
\Delta_\tau
- \frac{2}{11} \langle G^2\rangle_{T=0} \ , \nonumber \\ 
\frac{\alpha_s}{\pi} \langle {\cal B}^2\rangle_T &=& -\frac{4}{11} T^4 
\Delta_\sigma
+ \frac{2}{11} \langle G^2\rangle_{T=0} \ . \label{plaq}
\end{eqnarray}
The plaquette expectation values are related to the gluon condensate above
$T_c$ by~\cite{boyd,leut}
\begin{equation}
\langle G^2 \rangle_T = \langle G^2 \rangle_{T=0} -\Delta T^4 \ , 
\label{gconds}
\end{equation} 
where the interaction measure $\Delta = \Delta_\sigma +\Delta_\tau$ and 
$\langle G^2\rangle_{T=0} = (2.5\pm 1.0)T_c^4$.

The dispersion relation of a quark interacting with the thermal gluon
condensate is given by the roots of $D_\pm (L)=0$ (\ref{cond}).
The functions $a$ and $b$ (\ref{ab}) in (\ref{cond}) have been determined by
using (\ref{plaq}), where we took the plaquette expectation values from
the lattice calculations of Ref.~\cite{boyd}.
In Figs.1-3 we have displayed the dispersion relation of a quark having 
momentum $l$ for $T=1.1$ $T_c$, 2 $T_c$ and 4 $T_c$, respectively. As shown
in these figures there are two real positive solutions of $D_\pm(L)=0$.
For a given $T$ the upper curve $\omega_{+}(l)$ corresponds to the solution 
of $D_{+}(L)=0$, whereas the lower curve $\omega_{-}(l)$ represents the
solution of $D_{-}(L)=0$. Both branches are situated above the free
dispersion relation $\omega=l$, and start from a common effective
mass obtained in the $l\rightarrow 0$ limit as~\cite{markus}
\begin{equation}
\omega_{+}(0)=\omega_{-}(0)=m_{\rm{eff}}= \left [ \frac{2\pi\alpha_s}{3}
\left ( \langle {\cal E}^2\rangle_T +\langle {\cal B}^2\rangle_T 
\right )\right ]^{1/4} \ , \label{effm}
\end{equation}
which is given by $m_{\rm{eff}} \approx 1.15$ $T$ between $T=1.1$ $T_c$ 
and 4 $T_c$. The 
dispersion relation of a quark interacting with the in-medium gluon
condensate is similar to the dispersion relation obtained from the
hard thermal loop resummed (HTL) quark propagator \cite{ref3}. The 
$\omega_{+}(l)$ branch describes the propagation of an ordinary quark 
with thermal mass, and as in the HTL case we denote this 
quasiparticle by $q_{+}$, as
the ratio of its chirality to helicity is $+1$. 
For large momenta it is given by $\omega_+=l+c_1$, where $c_1$ is 
a constant containing the condensate \cite{markus}.
On the other hand, 
the $\omega_{-}(l)$ branch corresponds to the propagation of a quark mode
with a negative chirality to helicity ratio. This branch represents the 
plasmino
mode which is absent in the vacuum, and we denote this quasiparticle 
by $q_{-}$. As in the HTL case 
the $\omega_{-}(l)$ branch (plasmino mode) has a shallow minimum. 
For temperatures up to 2 $T_c$ this plasmino branch rapidly approaches 
the free dispersion relation (Figs.1 and 2). 
Owing to the strong increase of the magnetic condensate above
2 $T_c$ the plasmino branch stays clearly above the free dispersion 
relation for momenta under consideration
(see Fig.3). For high momenta, however, this branch 
approaches the free dispersion relation, $\omega_-\rightarrow l$. 
The plasmino mode
corresponds
to a purely collective long wave-length mode~\cite{ref3}. The 
residue of its pole
being proportional to $(\omega_{-}^2-l^2)^3$ for large momenta, becomes 
negligible for $l\gg T$. 
As a possible application of these quasiparticles
dispersion relations, we will study  the dilepton production from a
quark-gluon plasma. As we shall see below these dispersion relations 
will cause peaks in the dilepton production rate, which
could provide a possible signature of a quark-gluon plasma 
produced in heavy ion collisions.

\section{Dilepton production}

The dilepton production rate calculated from the photon self energy 
in the case of one massless lepton flavor is given by 
\begin{equation}
\frac{{\rm d}R}{{\rm d}^4x {\rm d}^4P} = -\frac{1}{12\pi^4} \frac{\alpha}
{M^2} \frac{1}{e^{E/T} - 1} {\rm{Im}}\Pi^\mu_\mu (P) , \label{rate}
\end{equation}
where $E=\sqrt {p^2 + M^2}$ is the energy of the photon with invariant 
mass $M$. 

Within the one-loop approximation the photon self energy (Fig.4) can be 
written using the Matsubara technique as
\begin{equation}
\Pi^{\mu\nu}(P) = - \frac{10}{3} e^2 T\sum_{k_0} \int 
\frac{{\rm d}^3k}{(2\pi)^3} {{\rm Tr}} \left [ \tilde S(K)\gamma^\mu
\tilde S(Q) \gamma^\nu \right ] , \label{impimunu}
\end{equation}
where $K$ and $Q=P-K$ are the fermionic loop four-momenta. In the following
only massless $u$ and $d$ quarks are considered.
We want to study the effect of an in-medium gluon condensate on the 
production rate of lepton pairs. This effect can be included by using
effective propagators containing the gluon condensate for the exchanged quarks.
In view of this we have replaced the bare propagators by the effective
propagators\footnote{The additional use of an effectice quark-photon
vertex will be discussed below.} 
in (\ref{impimunu}) as indicated by the filled circles in Fig.4.

Substitution of (\ref{effprop}) in (\ref{impimunu}) and performing the
traces yields~\cite{kapus}
\begin{eqnarray}
{\Pi}_\mu^\mu (P) &=& \frac{10}{3} e^2 T \sum_{k_0} \int \frac{{\rm d}^3k}
{(2\pi)^3}
\left [ \frac{1}{D_{+}(K)}\left ( \frac{1-\hat k \cdot \hat q}{D_{+}(Q)} +
\frac{1+\hat k \cdot \hat q}{D_{-}(Q)} \right ) \right. \nonumber \\
&& + \left. \frac{1}{D_{-}(K)}\left ( \frac{1+\hat k \cdot \hat q}{D_{+}(Q)} +
\frac{1-\hat k \cdot \hat q}{D_{-}(Q)} \right ) \right ] \ . \label{trimpi}
\end{eqnarray}
Now the imaginary part of the self energy can be computed by introducing 
spectral functions for the effective quark propagators as in the case of 
the real photon rate~\cite{kapus}. Following this work the imaginary part
of (\ref{trimpi}) can be expressed as
\begin{eqnarray}
{\rm{Im}} \Pi_\mu^\mu(P) &=& -\frac{10\pi}{3} e^2 \left (e^{E/T}-1 \right )
\int \frac {{\rm d}^3 k}{(2\pi)^3} \int_{-\infty}^\infty {\rm d}\omega  
\int_{-\infty}^\infty {\rm d}\omega' \ \nonumber \\
&& \times \delta \left ( E- \omega - \omega' \right ) 
n_F(\omega) n_F(\omega') \nonumber \\
&& \times \left [ \left (1+ {\hat q}\cdot 
{\hat k}\right ) \left \{ \rho_{+} \left ( \omega, k \right ) \rho_{-}
\left ( \omega', q \right ) 
+\rho_{-} \left ( \omega, k \right ) \rho_{+}\left ( \omega', q 
\right ) \right \} \right. \nonumber \\
&& \left. + \left (1- {\hat q}\cdot
{\hat k}\right ) \left \{ \rho_{+} \left ( \omega, k \right ) \rho_{+}
\left ( \omega', q \right )
+\rho_{-} \left ( \omega, k \right ) \rho_{-}\left ( \omega', q
\right ) \right \} \right ], \label{impi}
\end{eqnarray}
where $n_F$ is the Fermi-Dirac distribution and $\rho_\pm$ are the
spectral functions corresponding to $1/D_\pm (L)$ given by 
\begin{equation}
\rho_\pm (\omega, l) = R_\pm (\omega , l) \delta \left ( \omega - \omega_\pm 
\right ) + R_\mp (-\omega , l) \delta \left ( \omega + \omega_\mp \right ) \ ,
\label{spec}  
\end{equation}
where
\begin{equation}
R_\pm= \left | \frac{\left (\omega^2 -l^2\right )^3} {C_\pm}\right |  
\label{res}
\end{equation}
with
\begin{eqnarray}
C_\pm &=& - \left [ \left (1+a\right )\left (\omega^2-l^2\right )^3 
\ +\ \frac {b \left (\omega^2-l^2\right )^3}{\omega}
\ +\ 6\omega (\omega \mp l) (\omega^2-l^2)^2 \right. \nonumber \\ 
&& \left.  +\ \frac{g^2}{3} \omega (\omega \mp l) \left (\frac{5}{3} 
\langle {\cal E}^2\rangle_T
-\langle {\cal B}^2\rangle_T \right ) \ - \ \frac{8}{9}g^2\omega^2 
\langle {\cal E}^2\rangle_T
 \right ] . \label{resd}
\end{eqnarray}
The spectral functions in (\ref{spec}) contain only contributions from
the poles of the effective propagator corresponding to
the solutions $\omega_\pm$ of the dispersion relation 
$D_\pm(L)=0$ of the collective quark modes.
They do not have a 
contribution from discontinuities, corresponding 
to Landau damping \cite{pisarski},
because the effective quark propagator (\ref{effprop}) does not have an 
imaginary part coming from the quark self energy. 

Inserting (\ref{spec}) into (\ref{impi}) and performing the 
$\omega$-integrations, exploiting the delta functions 
of the spectral functions, one finds ($x=\hat p\cdot \hat k$)
\begin{eqnarray}
{\rm{Im}} \Pi_\mu^\mu(P) &=&  -\frac{5}{6\pi} e^2 \left (e^{E/T}-1 \right )
\int_0^\infty {\rm d}k\, k^2 \int_{-1}^{+1} {\rm d}x \nonumber \\
&& \ \ \  \times \left [ \left (1 \ + \ {\hat q}\cdot{\hat k}\right ) A \ + 
\ \left (  1 \ - \ {\hat q}\cdot{\hat k}\right ) B \right ], \ \label{impi1}
\end{eqnarray}
where
\begin{eqnarray}
A &=& n_F\left (\omega_{+}(k)\right )  n_F\left (\omega_{-}(q)\right )
R_{+}\left (\omega_{+}(k),k\right ) R_{-}\left (\omega_{-}(q),q\right )
\delta \left (E-\omega_{+}(k) -\omega_{-}(q) \right ) \nonumber \\
&+& n_F\left (-\omega_{-}(k)\right )  n_F\left (\omega_{-}(q)\right )
R_{-}\left (\omega_{-}(k),k\right ) R_{-}\left (\omega_{-}(q),q\right )
\delta \left (E+\omega_{-}(k) -\omega_{-}(q) \right ) \nonumber \\
&+& n_F\left (\omega_{+}(k)\right )  n_F\left (-\omega_{+}(q)\right )
R_{+}\left (\omega_{+}(k),k\right ) R_{+}\left (\omega_{+}(q),q\right )
\delta \left (E-\omega_{+}(k) +\omega_{+}(q) \right ) \nonumber \\
&+& n_F\left (-\omega_{-}(k)\right )  n_F\left (-\omega_{+}(q)\right )
R_{-}\left (\omega_{-}(k),k\right ) R_{+}\left (\omega_{+}(q),q\right )
\delta \left (E+\omega_{-}(k) +\omega_{+}(q) \right ) \nonumber \\
&+& n_F\left (\omega_{-}(k)\right )  n_F\left (\omega_{+}(q)\right )
R_{-}\left (\omega_{-}(k),k\right ) R_{+}\left (\omega_{+}(q),q\right )
\delta \left (E-\omega_{-}(k) -\omega_{+}(q) \right ) \nonumber \\
&+& n_F\left (-\omega_{+}(k)\right )  n_F\left (\omega_{+}(q)\right )
R_{+}\left (\omega_{+}(k),k\right ) R_{+}\left (\omega_{+}(q),q\right )
\delta \left (E+\omega_{+}(k) -\omega_{+}(q) \right ) \nonumber \\
&+& n_F\left (\omega_{-}(k)\right )  n_F\left (-\omega_{-}(q)\right )
R_{-}\left (\omega_{-}(k),k\right ) R_{-}\left (\omega_{-}(q),q\right )
\delta \left (E-\omega_{-}(k) +\omega_{-}(q) \right ) \nonumber \\
&+& n_F\left (-\omega_{+}(k)\right )  n_F\left (-\omega_{-}(q)\right )
R_{+}\left (\omega_{+}(k),k\right ) R_{-}\left (\omega_{-}(q),q\right )
\delta \left (E+\omega_{+}(k) +\omega_{-}(q) \right ) \nonumber \\
\label{ca}
\end{eqnarray}
and
\begin{eqnarray}
B &=& n_F\left (\omega_{+}(k)\right )  n_F\left (\omega_{+}(q)\right )
R_{+}\left (\omega_{+}(k),k\right ) R_{+}\left (\omega_{+}(q),q\right )
\delta \left (E-\omega_{+}(k) -\omega_{+}(q) \right ) \nonumber \\
&+& n_F\left (-\omega_{-}(k)\right )  n_F\left (\omega_{+}(q)\right )
R_{-}\left (\omega_{-}(k),k\right ) R_{+}\left (\omega_{+}(q),q\right )
\delta \left (E+\omega_{-}(k) -\omega_{+}(q) \right ) \nonumber \\
&+& n_F\left (\omega_{+}(k)\right )  n_F\left (-\omega_{-}(q)\right )
R_{+}\left (\omega_{+}(k),k\right ) R_{-}\left (\omega_{-}(q),q\right )
\delta \left (E-\omega_{+}(k) +\omega_{-}(q) \right ) \nonumber \\
&+& n_F\left (-\omega_{-}(k)\right )  n_F\left (-\omega_{-}(q)\right )
R_{-}\left (\omega_{-}(k),k\right ) R_{-}\left (\omega_{-}(q),q\right )
\delta \left (E+\omega_{-}(k) +\omega_{-}(q) \right ) \nonumber \\
&+& n_F\left (\omega_{-}(k)\right )  n_F\left (\omega_{-}(q)\right )
R_{-}\left (\omega_{-}(k),k\right ) R_{-}\left (\omega_{-}(q),q\right )
\delta \left (E-\omega_{-}(k) -\omega_{-}(q) \right ) \nonumber \\
&+& n_F\left (-\omega_{+}(k)\right )  n_F\left (\omega_{-}(q)\right )
R_{+}\left (\omega_{+}(k),k\right ) R_{-}\left (\omega_{-}(q),q\right )
\delta \left (E+\omega_{+}(k) -\omega_{-}(q) \right ) \nonumber \\
&+& n_F\left (\omega_{-}(k)\right )  n_F\left (-\omega_{+}(q)\right )
R_{-}\left (\omega_{-}(k),k\right ) R_{+}\left (\omega_{+}(q),q\right )
\delta \left (E-\omega_{-}(k) +\omega_{+}(q) \right ) \nonumber \\
&+& n_F\left (-\omega_{+}(k)\right )  n_F\left (-\omega_{+}(q)\right )
R_{+}\left (\omega_{+}(k),k\right ) R_{+}\left (\omega_{+}(q),q\right )
\delta \left (E+\omega_{+}(k) +\omega_{+}(q) \right ). \nonumber \\
\label{cb}
\end{eqnarray}

Changing the integration variable from $x$ to 
$q=|{\vec p}-{\vec k}| = \sqrt {p^2+k^2 -2pk x}$
the dilepton production rate (\ref{rate}) can be written as 
\begin{eqnarray}
\frac{{\rm d}R}{{\rm d}^4x {\rm d}^4P} &=& \frac{5}{36\pi^4}\frac 
{\alpha^2}{M^2} \frac{1}{p} 
\left [ \int_0^p  {\rm d}k \int_{p-k}^{p+k} {\rm d}q 
+ \int_p^\infty {\rm d}k \int_{k-p}^{p+k} {\rm d}q \right ] \nonumber \\
&& \times \left [ \left (p^2 - (k-q)^2 \right ) A + 
\left ( (k+q)^2 -p^2 \right ) B\right ] .\label{term1} 
\end{eqnarray}
Now one can perform the $q$-integration by means of the remaining 
$\delta$-functions in $A$ and $B$
leading to
\begin{eqnarray}
\frac{{\rm d}R}{{\rm d}^4x {\rm d}^4P} &=& \frac{5}{36\pi^4}\frac 
{\alpha^2}{M^2} \frac{1}{p} \int_0^\infty  {\rm d}k  
\left [\left (p^2 - (k-q_s)^2 \right ) 
\left ( A_1 + A_2 + A_3 + A_5 + A_6 + A_7 \right ) \right.\nonumber \\
&& \left. + \left ( (k+q_s)^2 -p^2 \right ) 
\left (B_1 + B_2 + B_3 + B_5 + B_6 +B_7 \right )\right ]_{|p-k|\le q_s \le 
p+k}\> , \label{term2} 
\end{eqnarray}
where the $q_s=q_s(E)$ determined by the various $\delta $-functions in (\ref{ca}) 
and (\ref{cb})
can assume two different values in the case of the plasmino branch due to the
presence of the minimum and
\begin{eqnarray}
A_1 &=& n_F\left (\omega_{+}(k)\right )  n_F\left (\omega_{-}(q_s)\right )
R_{+}\left (\omega_{+}(k),k\right ) 
\frac{R_{-}\left (\omega_{-}(q_s),q_s\right )}
{|{\rm d}\omega_{-}(q)/{\rm d}q|_{q_s}}, \nonumber \\
A_2 &=& n_F\left (-\omega_{-}(k)\right )  n_F\left (\omega_{-}(q_s)\right )
R_{-}\left (\omega_{-}(k),k\right ) 
\frac {R_{-}\left (\omega_{-}(q_s),q_s\right )}
{|{\rm d}\omega_{-}(q)/{\rm d}q|_{q_s}}, \nonumber \\
A_3 &=& n_F\left (\omega_{+}(k)\right )  n_F\left (-\omega_{+}(q_s)\right )
R_{+}\left (\omega_{+}(k),k\right ) 
\frac {R_{+}\left (\omega_{+}(q_s),q_s\right )}
{|{\rm d}\omega_{+}(q)/{\rm d}q|_{q_s}}, \nonumber \\
A_5&=& n_F\left (\omega_{-}(k)\right )  n_F\left (\omega_{+}(q_s)\right )
R_{-}\left (\omega_{-}(k),k\right ) 
\frac{R_{+}\left (\omega_{+}(q_s),q_s\right )}
{|{\rm d}\omega_{+}(q)/{\rm d}q|_{q_s}}, \nonumber \\
A_6&=& n_F\left (-\omega_{+}(k)\right )  n_F\left (\omega_{+}(q_s)\right )
R_{+}\left (\omega_{+}(k),k\right ) 
\frac{R_{+}\left (\omega_{+}(q_s),q_s\right )}
{|{\rm d}\omega_{+}(q)/{\rm d}q|_{q_s}}, \nonumber \\
A_7&=& n_F\left (\omega_{-}(k)\right )  n_F\left (-\omega_{-}(q_s)\right )
R_{-}\left (\omega_{-}(k),k\right ) 
\frac{R_{-}\left (\omega_{-}(q_s),q_s\right )}
{|{\rm d}\omega_{-}(q)/{\rm d}q|_{q_s}}, \nonumber \\
B_1 &=& n_F\left (\omega_{+}(k)\right )  n_F\left (\omega_{+}(q_s)\right )
R_{+}\left (\omega_{+}(k),k\right ) 
\frac {R_{+}\left (\omega_{+}(q_s),q_s\right )}
{|{\rm d}\omega_{+}(q)/{\rm d}q|_{q_s}}, \nonumber \\
B_2&=& n_F\left (-\omega_{-}(k)\right )  n_F\left (\omega_{+}(q_s)\right )
R_{-}\left (\omega_{-}(k),k\right ) 
\frac{R_{+}\left (\omega_{+}(q_s),q_s\right )}
{|{\rm d}\omega_{+}(q)/{\rm d}q|_{q_s}}, \nonumber \\
B_3&=& n_F\left (\omega_{+}(k)\right )  n_F\left (-\omega_{-}(q_s)\right )
R_{+}\left (\omega_{+}(k),k\right ) 
\frac{R_{-}\left (\omega_{-}(q_s),q_s\right )}
{|{\rm d}\omega_{-}(q)/{\rm d}q|_{q_s}}, \nonumber \\
B_5&=& n_F\left (\omega_{-}(k)\right )  n_F\left (\omega_{-}(q_s)\right )
R_{-}\left (\omega_{-}(k),k\right ) 
\frac{R_{-}\left (\omega_{-}(q_s),q_s\right )}
{|{\rm d}\omega_{-}(q)/{\rm d}q|_{q_s}}, \nonumber \\
B_6&=& n_F\left (-\omega_{+}(k)\right )  n_F\left (\omega_{-}(q_s)\right )
R_{+}\left (\omega_{+}(k),k\right ) 
\frac{R_{-}\left (\omega_{-}(q_s),q_s\right )}
{|{\rm d}\omega_{-}(q)/{\rm d}q|_{q_s}}, \nonumber \\
B_7&=& n_F\left (\omega_{-}(k)\right )  n_F\left (-\omega_{+}(q_s)\right )
R_{-}\left (\omega_{-}(k),k\right ) 
\frac{R_{+}\left (\omega_{+}(q_s),q_s\right )}
{|{\rm d}\omega_{+}(q)/{\rm d}q|_{q_s}}, \nonumber \\
\label{cabs}
\end{eqnarray}

The group velocity factors in (\ref{cabs}) follow from the dispersion
relation, $D_\pm(L)=0$, of (\ref{cond}) as 
\begin{equation}
\frac{{\rm d}\omega_\pm(l)}{{\rm d}l}
= \pm \frac{F_\pm \left (\omega_{\pm}(q),a, b, l\right )} 
{G_\pm \left (\omega_{\pm}(q),a, b, l\right )} \ , \label{s}
\end{equation}
where
\begin{eqnarray}
F_\pm&=& \left (1+a\right )^2 \left (\omega_{\pm}^2(l)-l^2 \right )^3 
\ \mp \ 6b \left (\omega_{\pm}^2(l)-l^2 \right )^2 l 
\ \mp \ \frac{g^2}{6} b \left (\frac{2}{3}\langle {\cal E}^2\rangle_T 
-\frac{2} {5}\langle {\cal B}^2\rangle_T \right )l \nonumber \\
&& \pm \ \frac{8}{45}g^2 \left (1+a\right ) l \omega_{\pm}(l) 
\langle {\cal B}^2\rangle_T \ \ , \nonumber \\
G_\pm&=& -\left (1+a\right )C_\pm \ ,
  \label{fg}
\end{eqnarray}
and $a$, $b$, and $C_\pm$ are given in (\ref{ab}) and
(\ref{resd}), respectively. As we will see in the next section
the group velocity leads to a characteristic feature of the dilepton rate.

In (\ref{term2}) we have dropped terms $A_4$,
$A_8$, $B_4$ and $B_8$ as the corresponding $\delta$-functions in (\ref{cb})
can never be satisfied by virtue of energy conservation
since $\omega_\pm$ is always positive. Now, one
can perform the $k$-integration in (\ref{term2}) numerically, and we find 
that the terms, which satisfy the energy conservation, correspond to 
various physical processes involving two quasiparticles with different
momentum $k$ and $q$. 

However, 
before discussing our results, we would like to give the corresponding
dilepton production rate for ${\vec p}=0$. For this purpose we combine
(\ref{rate}) and (\ref{impi1}) with $\vec q = -\vec k$ and obtain
\begin{eqnarray}
\frac{{\rm d}R}{{\rm d}^4x {\rm d}^4P}(\vec p=0) &=& \frac{10}{9\pi^4}\frac 
{\alpha^2}{M^2} \int_0^\infty {\rm d}k\, k^2 
 \left [ n_F^2\left (\omega_{+}(k)\right ) R_{+}^2\left (\omega_{+}(k)\right) 
\delta\left (E-2\omega_{+}(k)\right ) \right. \nonumber \\ 
&+& \left. 2n_F\left (\omega_{+}(k)\right )n_F\left (-\omega_{-}(k)\right )
R_{+}\left (\omega_{+}(k)\right ) R_{-}\left (\omega_{-}(k)\right )
\delta\left (E-\omega_{+}(k)+ \omega_{-}(k)\right ) \right. \nonumber \\ 
&+& \left. 2n_F\left (\omega_{-}(k)\right )n_F\left (-\omega_{+}(k)\right )
R_{+}\left (\omega_{+}(k)\right ) R_{-}\left (\omega_{-}(k)\right )
\delta\left (E+\omega_{+}(k)- \omega_{-}(k)\right ) \right. \nonumber \\   
&+& \left. n_F^2\left (\omega_{-}(k)\right ) R_{-}^2\left (\omega_{-}(k)\right) 
\delta\left (E-2\omega_{-}(k)\right ) \right ] \ .   \label{ratep0}
\end{eqnarray}    
After performing the $k$-integration by means of the $\delta$-functions, the
expression for the dilepton rate at $\vec p=0$ becomes
\begin{eqnarray}
\frac{{\rm d}R}{{\rm d}^4x {\rm d}^4P} (\vec p=0) &=& \frac{10}{9\pi^4}\frac 
{\alpha^2}{M^2}  \sum_{k_s} k_s^2  
\left [ n_F^2\left (\omega_{+}(k_s)\right ) 
R_{+}^2\left (\omega_{+}(k_s)\right)
\frac{1}{2}\left |\frac{{\rm d}\omega_{+}(k)}{{\rm d}k}\right |_{k_s}^{-1} 
\right. \nonumber \\
&+& \left. 2n_F\left (\omega_{+}(k_s)\right )n_F\left (-\omega_{-}(k_s)\right )
R_{+}\left (\omega_{+}(k_s)\right ) R_{-}\left (\omega_{-}(k_s)\right )
\left |\frac{{\rm d}\left (\omega_{+}(k)-\omega_{-}(k)\right )}
{{\rm d}k}\right |_{k_s}^{-1}
\right. \nonumber \\ 
&+& \left. 2n_F\left (\omega_{-}(k_s)\right )n_F\left (-\omega_{+}(k_s)\right )
R_{+}\left (\omega_{+}(k_s)\right ) R_{-}\left (\omega_{-}(k_s)\right )
\left | \frac{{\rm d}\left (\omega_{-}(k)-\omega_{+}(k)\right )}
{{\rm d}k}\right |_{k_s}^{-1}
 \right. \nonumber \\   
&+& \left. n_F^2\left (\omega_{-}(k_s)\right ) 
R_{-}^2\left (\omega_{-}(k_s)\right) 
\frac{1}{2} \left |\frac{{\rm d}\omega_{-}(k)}{{\rm d}k}\right |_{k_s}^{-1} 
\right ] \ .   \label{ratep1}
\end{eqnarray}    

\section{\bf Results and Discussion}

First we would like to discuss the dilepton production from a 
quark-gluon plasma at momentum $\vec p=0$ of the virtual photon. 
The corresponding rate is given by
(\ref{ratep1}).  The different terms in (\ref{ratep1}) correspond to
various physical processes involving two quasiparticles $q_+$ and $q_-$ 
with same momentum $k$. 
The first term represents the annihilation process  $q_+\bar q_+\rightarrow
\gamma^*$. The second term corresponds to $q_+\rightarrow q_-\gamma^*$, a decay
process from a $q_+$-mode to a plasmino plus a virtual photon. 
Energy 
conservation does not allow the process given by the third term
($q_-\rightarrow \bar q_+\gamma^*$).
Finally, the fourth term corresponds to a process, 
$q_-\bar q_- \rightarrow \gamma^*$, i.e. annihilation of
plasmino modes. The static differential rate of the aforementioned 
processes are displayed in Fig.5 for $T=1.1$ $T_c$ (solid line), 2 $T_c$
(dashed curve) and 4 $T_c$ (dotted curve). Similar to the  
HTL case \cite{ref3}
the partial rate in the presence of a 
gluon condensate shows peaks (van Hove 
singularities) at different invariant masses of the virtual photon. Below we 
discuss the contributions to the rate from each process in detail.

The channel, $q_+\rightarrow q_-\gamma^*$, opens up at $M=0$. 
This process continues up to $M=1.01$ $T_c$ for 
$T=1.1$ $T_c$, $M=1.83$ $T_c$ for $T=2$ $T_c$ and $M=2.14$ $T_c$ for $T=4$ $T_c$,
respectively, where the first peak appears due to the vanishing
group velocity $dE/dk=0$ at the maximum $E=M=\omega_{+}(k)-\omega_-(k)$,
since the density of states, which enters the rate (\ref{ratep1}), is 
inversely proportional to the group velocity. 

The $q_+\rightarrow q_-\gamma^*$ channel terminates at the peak, after which 
there is a gap because neither of the other processes is possible in this
invariant mass regime. The size of the gap depends on the
temperature. For $T=1.1$ $T_c$ it ranges from $M=1.01$ $T_c$ to 2.07 $T_c$, for 
$T=2$ $T_c$ from $M=1.83$ $T_c$ to 3.73 $T_c$, and for $T=4$ $T_c$  from $M=2.14$ $T_c$ to 
8.76 $T_c$.
  
The process, $q_-\bar q_-\rightarrow \gamma^*$, starts at an energy which is 
twice the energy of the minimum of the plasmino branch,
$E=M=2\omega_{-}(k_{min})$. The diverging density of states at that point
again causes a van Hove singularity.
This process continues with increasing $M$ but
falls off very fast due to two reasons:
i) as $M$ increases the high energy plasmino modes come into the
game and the corresponding square of the residue $R_-^2(\omega_-(k),k)$,
to which the rate is proportional, becomes very
small since it is proportional to $(\omega_-^2(k)-k^2)^3$, and
ii) with increasing $M$ the density of states decreases gradually.

At $M=E=2\omega_{+}(k)\ge 2m_{\rm{eff}}$, the process, 
$q_{+}\bar q_+\rightarrow \gamma^*$, shows up. As $M$ increases, the
contribution from this process grows and dominates over the plasmino 
annihilation 
process, resulting in a dip in the dilepton rate. For large $M$ this 
annihilation process is solely responsible for the dilepton rate. 

For $T=4$ $T_c$ the
contribution from the plasmino annihilation is nearly as big as the 
$q_{+}$-$\bar q_+$-annihilation at the mass regime displayed in Fig.4
because of the fairly large deviation of the plasmino branch from
the free dispersion relation. This is the reason why the dilepton rate
is almost flat at large masses in Fig.5. Only at very large invariant masses
the contribution from $q_{-}$-$\bar q_-$-annihilation will vanish.

The channel $q_{+}\bar q_-\rightarrow \gamma^*$, which contributes in the 
case of the HTL calculation \cite{ref3}, is absent here. This can be traced 
back to the fact that we did not take into account an effective quark-photon 
vertex. Using such a vertex, (\ref{trimpi}) does not hold anymore and terms 
containing $\delta (E-\omega _+(k)-\omega _-(k))$ show up in (\ref{ratep0}).

Comparing our rate to the HTL result \cite{ref3} for realistic values of 
the strong coupling constant appearing in the HTL rate, e.g. $g=2.5$, we 
observe
that our rate is smaller by about a factor of 5 to 10 than the pole-pole 
contribution of the HTL calculation. 
At large $M$ the HTL rate reduces to the Born rate \cite{ref3,xxx}
\begin{equation}
\frac{{\rm d}R^{\rm{Born}}}{{\rm d}^4x {\rm d}^4P} (\vec p=0) 
= \frac{5\alpha^2}{36\pi^4} \> \left (e^{M/2T}+1\right )^{-2}  \ , 
\label{born} 
\end{equation}
which follows from the first term of (\ref{ratep0}) in the limit 
$\omega_+ \rightarrow k$ where $R_+^{HTL} \rightarrow 1$. In contrast,
the rate given here does not reduce to the Born rate in the large $M$
limit, since $R_+$ is given by $1/4$ instead of 1 for infinite 
large momenta. This can be seen easily from (\ref{res}) using 
the asymptotic form of the $\omega_+$ branch, $\omega_0^+\rightarrow
k+c_1$, discussed in section 2. The fact that $R_+$, which enters the rate 
(\ref{ratep0}) quadratically, is always small causes the suppression
of the rate compared to the HTL case. Another reason for our 
small rate might be the fact that we did not take into account
an effective quark-photon vertex in our calculation. Neglecting this
vertex in the HTL computation leads to a reduction of the dilepton
rate by a factor 2 to 10 depending on $M$.

Furthermore, since the spectral densities
of the effective quark propagator have no discontinuous parts there is no 
contribution from cuts as opposed to \cite{ref3}. In the HTL case the cut 
contribution, which shows no dramatic structures, dominates over the 
pole-pole contribution and covers the peaks and gap of the latter completely.

Finally we turn to the dilepton rate at non-zero virtual photon
momentum. The corresponding rate is given in (\ref{term2}). 
The processes corresponding to terms $A_2$, $A_3$, 
$A_6$, $A_7$, $B_6$ and $B_7$, namely transitions within a branch and 
transitions from the lower to the upper branch, do not contribute to 
the rate, because they are forbidden for timelike photons decaying into
dileptons due to energy conservation \cite{ref4}.
The processes corresponding to $A_1$ and $A_5$ indicate annihilation 
between a quark ($q_+$) and a plasmino mode ($q_-$) with different momentum
to a virtual photon with energy $E$, which were absent at $p=0$. The
process given by $B_1$ is the annihilation between a quark and antiquark
($q_+(k)\bar q_+(q)\rightarrow \gamma^*$), whereas $B_5$ corresponds to 
the annihilation 
($q_-(k)\bar q_-(q)\rightarrow \gamma^*$) between two plasmino modes. 
The term $B_2$ corresponds to the
decay process, $q_+ (q)\rightarrow q_-(k)\gamma^*$, whereas $B_3$ to
$q_+(k)\rightarrow q_-(q)\gamma^*$.
The differential rate involving these processes
are displayed in Figs.6-8 for virtual photon momenta $p=1$ $T_c$, 2 $T_c$, and
3 $T_c$, respectively, at different temperatures, namely $T=1.1$ $T_c$ 
(solid line), 2 $T_c$ (dashed line), and 4 $T_c$ (dotted line).

For $p=1$ $T_c$ (Fig.6) the decay processes corresponding to terms 
$B_2$ and $B_3$ open up at $E(=\sqrt{p^2+M^2})$=1 $T_c$. 
These processes continue up to $E=2.0$ $T_c$ for $T=1.1$ $T_c$, $E=2.8$ $T_c$ for 
$T=2$ $T_c$, and $E=3.3$ $T_c$ for $T=4$ $T_c$, respectively.
The first peak at $p=0$ is smeared out and a shoulder appears, 
which becomes more and more significant as temperature 
increases. At very low invariant masses the contribution from these 
decay processes to the rate is large caused by a very large density of
states due to kinematical reasons. With increasing
$E$ the density of states decreases and at the same time 
the residue of the
plasmino mode falls off gradually. The interplay of these two quantities 
is responsible for the shoulder in the rate, which is more pronounced at higher
$T$. 

The process corresponding to the annihilation of two plasmino modes 
(term $B_5$) turns on at $E=2.1$ $T_c$ for $T=1.1$ $T_c$, 
$E=3.6$ $T_c$ for $T=2$ $T_c$, and $E=8.7 $ $T_c$ for $T=4$ $T_c$, 
respectively, 
with a smoothened van Hove peak and falls off rapidly due to the reasons 
explained above. The annihilation processes involving
a quark and a plasmino mode (terms $A_1$ and $A_5$) open up 
at $E=2.4$ $T_c$ for $T=1.1$ $T_c$, $E=4.1$ $T_c$ for $T=2$ $T_c$, and
$E=9.1$ $T_c$ for $T=4$ $T_c$, respectively. With increasing $E$ 
the contribution from this two 
processes falls off rapidly. As a result of the interplay of these processes
a small bump or plateau appears in the dilepton production rate. 
The annihilation process involving 
a $q_+$ and a $\bar q_+$ (term $B_1$) opens up at $E=3.1$ $T_c$ for $T=1.1$ $T_c$, 
$E=5.2$ $T_c$ for $T=2$ $T_c$, and $E=9.7$ $T_c$ for $T=4$ $T_c$.
As explained for $p=0$, the contribution from this process
grows and dominates over the other processes at large $E$. 
The dip and the second bumb
for lower $T$ come from the interplay of the decreasing processes involving
plasminos and the increasing process involving only quark modes.
For higher $T$ these structures vanish as in the $p=0$ case.

In Fig.7 we have displayed the results for $p=2$ $T_c$. 
The gap for $T=4$ $T_c$ becomes narrower whereas the gap disappears 
for $T=1.1$ $T_c$ and 2 $T_c$. For all $T$ the decay processes 
begin at  $E=2$ $T_c$ and continue up to
$E=3.0$ $T_c$ for $T=1.1$ $T_c$, $E=3.9$ $T_c$ for $T=2$ $T_c$, and $E=4.6$ $T_c$ for 
$T=4$ $T_c$. For $T=1.1$ $T_c$ the annihilation process involving two 
plasmino modes opens up around 
$E=2.3$ $T_c$ and for $T=2$ $T_c$ at $E=3.7$ $T_c$, causing 
the gap to disappear. The process involving annihilation
of a quark and a plasmino mode opens up around $E=3.2$ $T_c$ for $T=1.1$ $T_c$,
leading to a deep dip in the dilepton rate. As $E$
increases the rate falls off and is finally dominated gradually by the  
quark-antiquark annihilation process, causing a second dip
around $E=4.3$ $T_c$. For $T=2$ $T_c$ the small kink around $E=4$ $T_c$ 
is due to the interplay of decay and annihilation processes involving
plasminos. At this temperature 
the annihilation process between a quark and a plasmino 
mode becomes active and results in a dip
at $E\simeq 4.5$ $T_c$.
Again the second dip around $E=6.2$ $T_c$ is due to the interplay of
the usual annihilation process and annihilation processes involving 
plasminos. The overall feature for $T=4$ $T_c$ 
remains the same as that at $p=1$ $T_c$. 

As a last example the partial dilepton production rate for $p=3$ $T_c$ is 
displayed in Fig.8 for
$T=(1.1-4)$ $T_c$. The features for $T=1.1$ $T_c$ are the same as at $p=2$
$T_c$ except that
the different processes involved now open up at higher energies.
For $T=2$ $T_c$ the quark-plasmino annihilation processes starts at
around $E=5.3$ $T_c$ in such a way that the valley like shape in the rate
turns into a deep pocket. As expected, the gap width for
$T=4$ $T_c$ shrinks a bit. Now the dilepton rate immediately after the gap 
is rather flat as the contribution from annihilation process 
involving two plasmino modes has stabilized. At
$E=8.8$ $T_c$ the step like structure in the rate is caused
by the onset of the annihilation processes between a quark and a plasmino mode.
The usual annihilation process begins only
around $E=11$ $T_c$, which is not shown in the figure.

Finally we want to discuss the neglect of
an effective quark-photon vertex and possible observable consequences
of our results. Such a vertex, 
related by the Ward identity
to the effective quark propagator, should be taken into account when
calculating the photon self energy. As a matter of fact, in the case
of $\pi^+$-$\pi^-$-annihilation the consideration of the Ward identity
lead to a strong suppression of the dilepton production \cite{ref11}. 
In contrast, taking into account the effective quark-photon vertex in 
the HTL calculation leads to an increase of the rate by a factor 2 to
10 depending on $M$. Unfortunately,
the computation of the effective quark-photon vertex and the photon self energy
with such a vertex will be very involved. However, the positions of
the singularities and the gap is solely determined by the quark
dispersion relation. Our main result is that non-perturbative quark dispersion
relations in the QGP lead to sharp structures in the dilepton rate at
the positions shown in Figs.5-8.

The absolute value of the rate is determined not only by the consideration of the
effective quark-photon vertex but also by higher order damping effects which will
broaden the peaks.
Whether these structures can be seen in the dilepton spectrum or
not depends on the amount of broadening of the peaks and 
on the smoothening of these structures by the space-time evolution of the 
fireball. Furthermore processes involving additional 
gluons such as Compton scattering \cite{ref3} 
and bremsstrahlung \cite{ref5}, which lead to a smooth dilepton rate \cite{ref3}, and hadronic
processes might cover 
up these structures. If, however, new structures will be observed in the low mass
($M<1$ GeV) dilepton spectrum, they will provide a strong indication for the presence
of deconfined, collective quarks in the QGP, in particular since the hadronic
contribution to the dilepton rate is
expected to be smooth due to medium effects \cite{ref12}. 
As a matter of fact, this argument does not 
depend on a specific approximation scheme \cite{ref13} as the general behavior (two branches,
common effective mass, plasmino minimum, free dispersion at large momenta) of the quark 
dispersion agrees with the one shown in Fig.1-3 and found in Ref.\cite{ref3}. Although 
such structures have not been observed so far
due to the small life time of the QGP phase at SPS (if it exists at all)
as indicated by hydrodynamical calculation \cite{ref14}, it will be worthwhile to
look for new structures in the low mass dilepton spectrum (in particular at low
photon momenta) at RHIC and LHC,
where the thermal dilepton spectrum is expected to be dominated by the
QGP phase.

\vspace*{1cm}
\centerline{\bf ACKNOWLEDGMENTS}
\vspace*{1cm}
We are grateful to J. Engels for useful correspondence.

\begin{figure}
\centerline{\psfig{figure=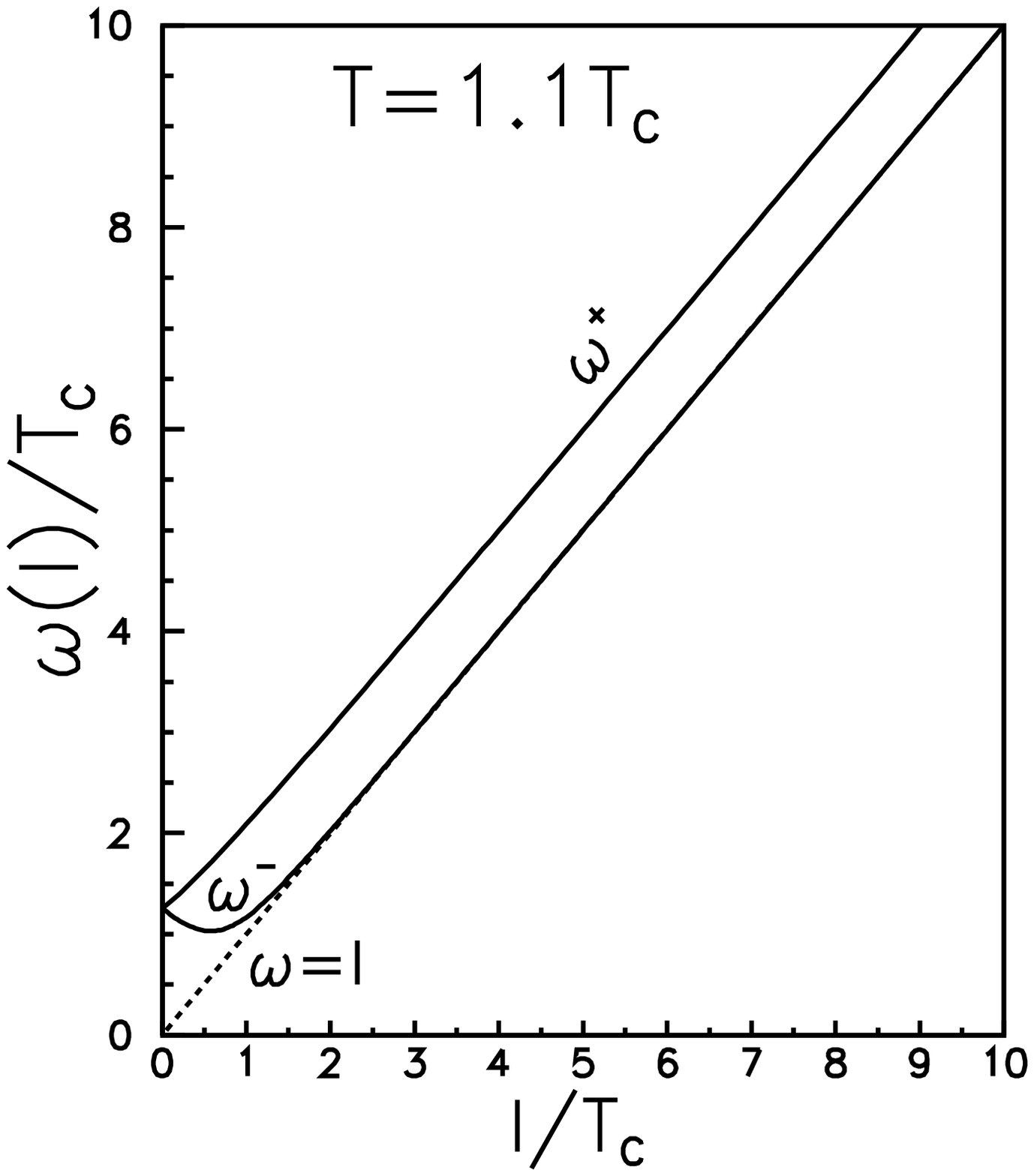,width=9cm}}
\vspace*{-1cm}
\caption{Quark dispersion relation $\omega(l)/T_c$ versus $l/T_c$ in 
a QGP at $T=1.1$ $T_c$ in the presence of a gluon condensate.}
\end{figure}

\begin{figure}
\centerline{\psfig{figure=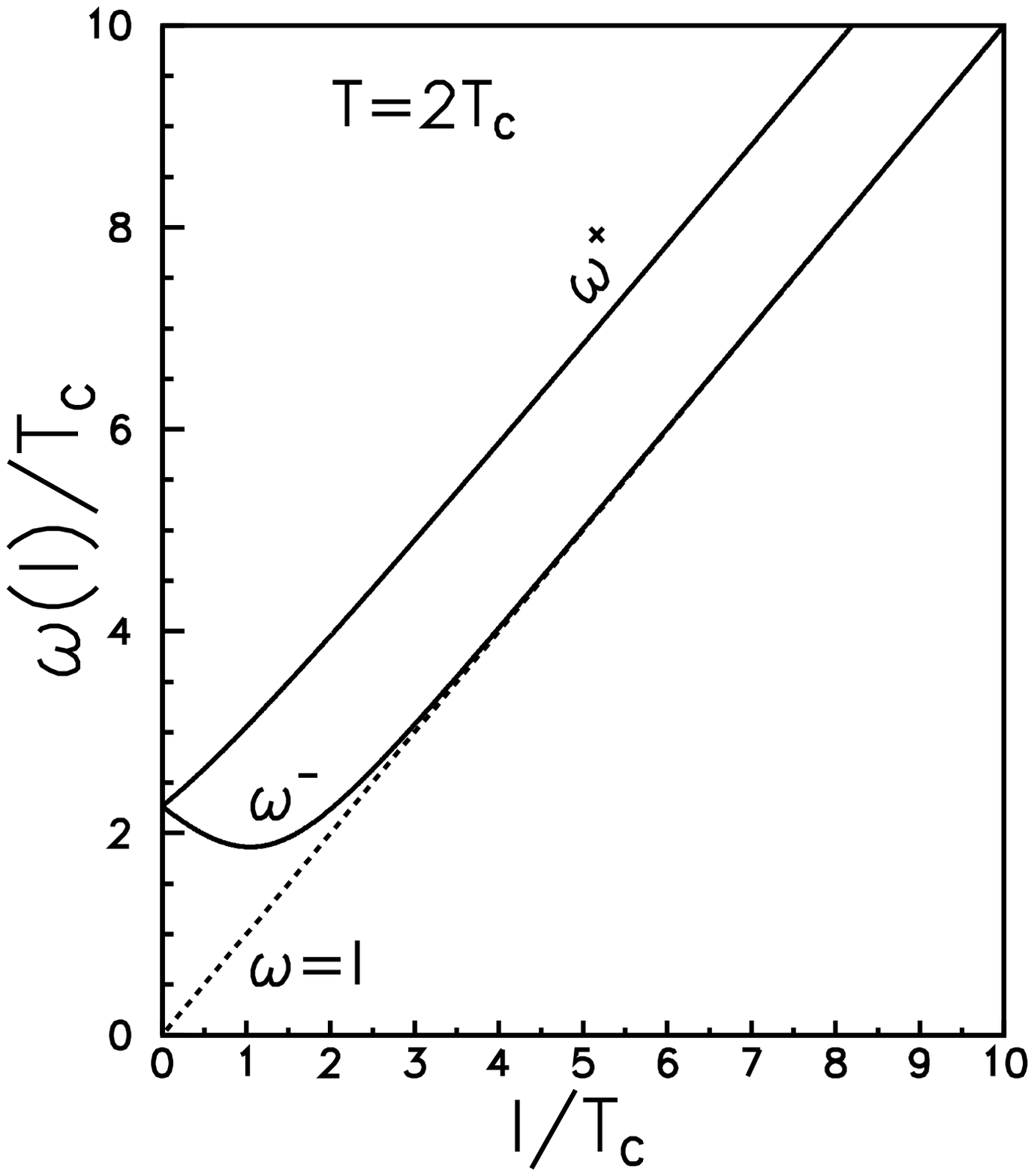,width=9cm}}
\vspace*{-1cm}
\caption{Quark dispersion relation $\omega(l)/T_c$ versus $l/T_c$ in 
a QGP at $T=2$ $T_c$ in the presence of a gluon condensate.}
\end{figure}

\begin{figure}
\centerline{\psfig{figure=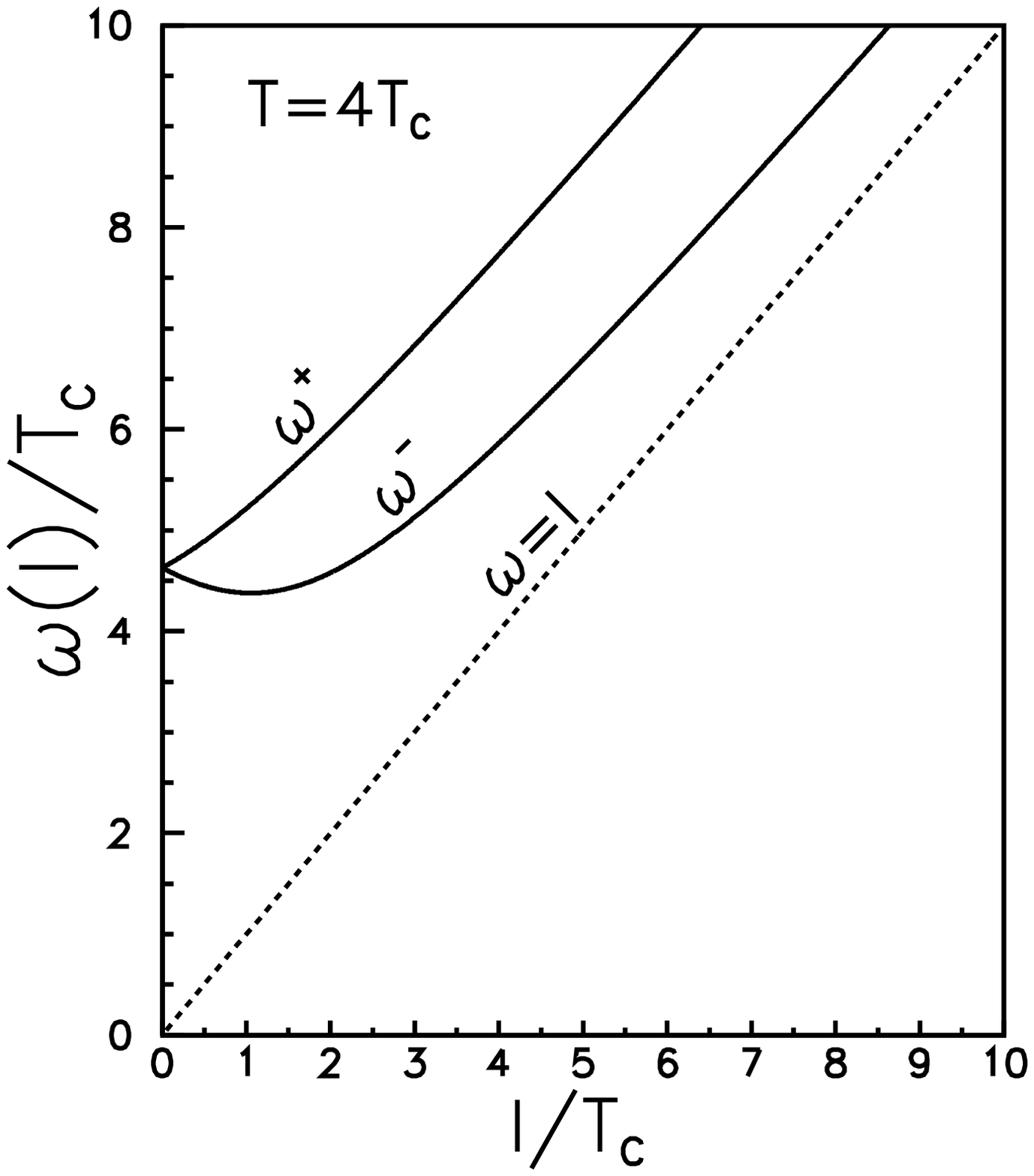,width=9cm}}
\vspace*{-1cm}
\caption{Quark dispersion relation $\omega(l)/T_c$ versus $l/T_c$ in 
a QGP at $T=4$ $T_c$ in the presence of a gluon condensate.}
\end{figure}

\begin{figure}
\centerline{\psfig{figure=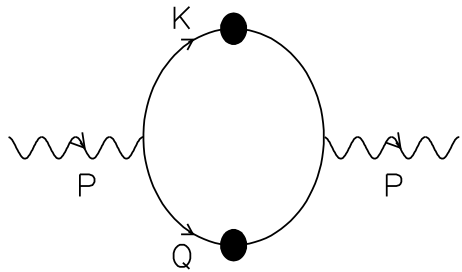,width=9cm}}
\vspace*{-1cm}
\caption{One-loop photon self energy with effective quark propagators
containing the gluon condensate}
\end{figure}

\begin{figure}
\centerline{\psfig{figure=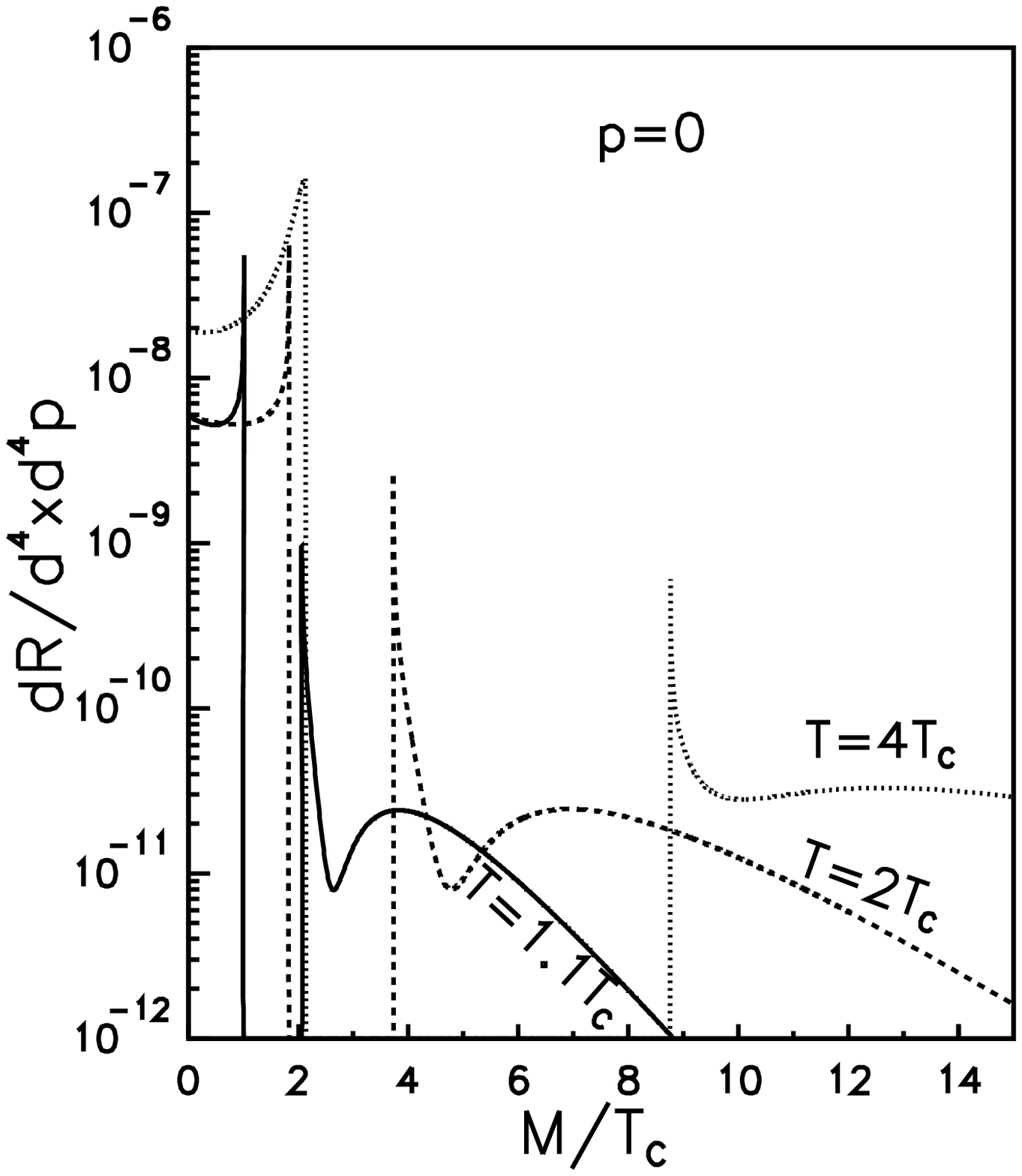,width=9cm}}
\vspace*{-1cm}
\caption{Dilepton production rate from a QGP in the presence of a gluon 
condensate at photon momentum $p=0$.}
\end{figure}

\begin{figure}
\centerline{\psfig{figure=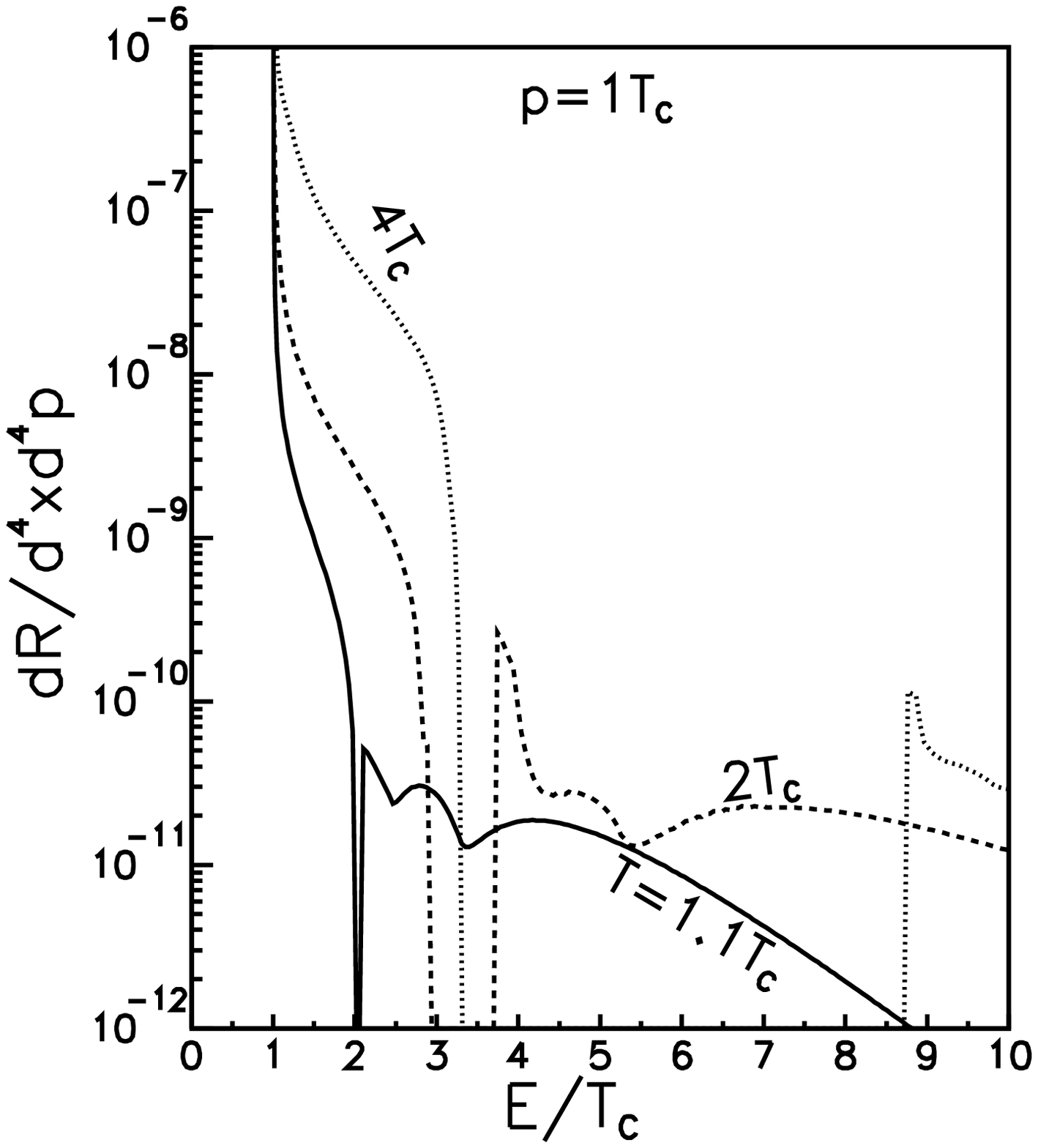,width=9cm}}
\vspace*{-1cm}
\caption{Dilepton production rate from a QGP in the presence of a gluon 
condensate at photon momentum $p=1$ $T_c$.}
\end{figure}

\begin{figure}
\centerline{\psfig{figure=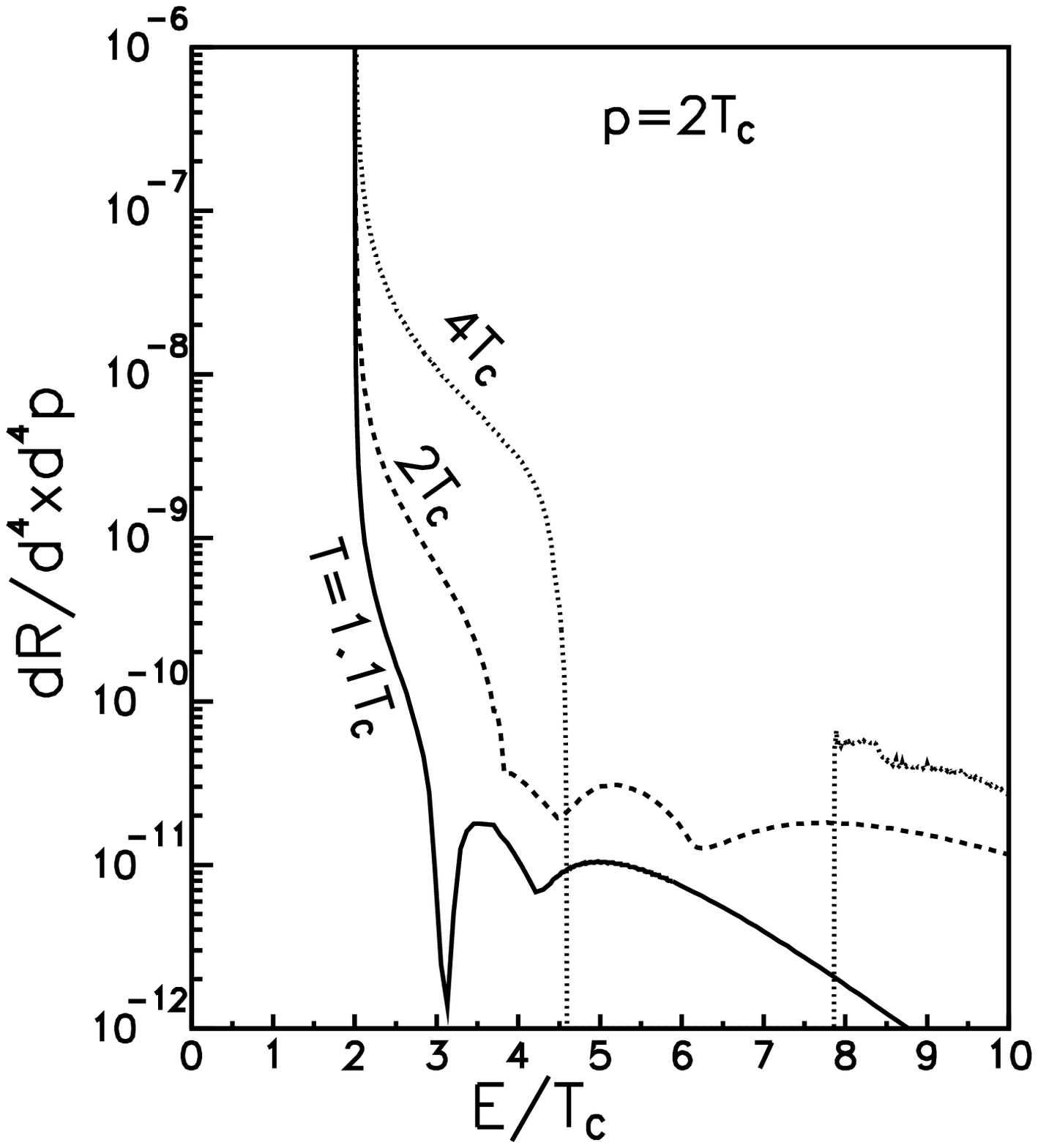,width=9cm}}
\vspace*{-1cm}
\caption{Dilepton production rate from a QGP in the presence of a gluon 
condensate at photon momentum $p=2$ $T_c$.}
\end{figure}

\begin{figure}
\centerline{\psfig{figure=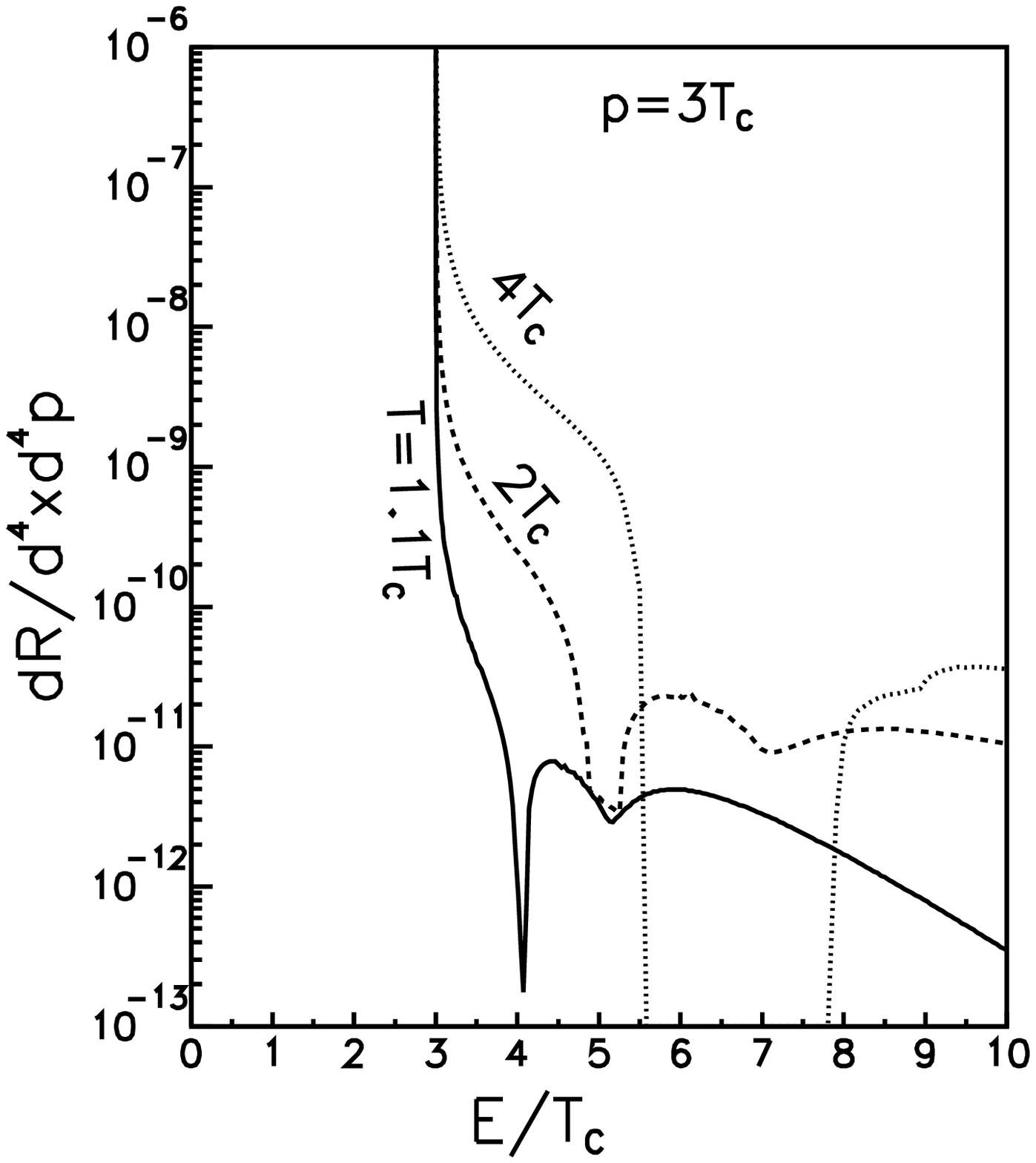,width=9cm}}
\vspace*{-1cm}
\caption{Dilepton production rate from a QGP in the presence of a gluon 
condensate at photon momentum $p=3$ $T_c$.}
\end{figure}


\begin{references}
\bibitem{ref1} P.V. Ruuskanen, Nucl. Phys. {\bf A544}, 169c (1992).
\bibitem{ref2} M.H. Thoma, Phys. Rev D {\bf 51}, 862 (1995). 
\bibitem{ref3} E. Braaten, R.D. Pisarski, and T.C. Yuan, Phys. Rev. Lett.
{\bf 64}, 2242 (1990).
\bibitem{ref4} S.M.H. Wong, Z. Phys. C {\bf 53}, 465 (1992).
\bibitem{ref5} P. Aurenche, F. Gelis, R. Kobes, and H. Zaraket, Phys.
Rev. D {\bf 58}, 085003 (1998); P. Aurenche, F. Gelis, R. Kobes, and H. 
Zaraket, hep-ph/9903307. 
\bibitem{ref6} A. Peshier, B. K\"ampfer, O.P. Pavlenko, and G. Soff, 
Phys. Lett. B {\bf 337}, 235 (1994); A. Peshier, B. K\"ampfer, O.P. Pavlenko, 
and G. Soff, Phys. Rev. D {\bf 54}, 2399 (1996).
\bibitem{ref7} G. Boyd, S. Gupta, F. Karsch, and E. Laermann, Z. Phys. C
{\bf 64}, 331 (1994).
\bibitem{boyd} G. Boyd et al., Nucl. Phys. {\bf B469}, 419 (1996).
\bibitem{markus} A. Sch{\"a}fer and M. H. Thoma, Phys. Lett. B {\bf 451}, 195 (1999).
\bibitem{ref10} I. Schmidt and J.J. Yang, hep-ph/9906510.
\bibitem{lavelle} M. J. Lavelle and M. Schaden, Phys. Lett. B {\bf 208}, 419 (1988).
\bibitem{leut} H. Leutwyler, in Proc. Conf. QCD - 20 years later, Eds. P. M.
Zerwas and H. A. Kastrup (World Scientific, Singapore,1993) p. 693.
\bibitem{kapus} J. I. Kapusta, P. Lichard, and D. Seibert, Phys. Rev. 
D {\bf 44}, 2774 (1991).
\bibitem{pisarski} R. D. Pisarski, Physica A {\bf 158}, 146 (1989).
\bibitem{xxx} J. Cleymans, J. Fingberg, and K. Redlich, Phys. Rev. D {\bf 35}, 2153 (1987).
\bibitem{ref11} C.L. Korpa and S. Pratt, Phys. Rev. Lett. {\bf 64}, 1502 (1991).
\bibitem{ref12} R. Rapp and J. Wambach, hep-ph/9907502.
\bibitem{ref13} A. Peshier and M.H. Thoma, hep-ph/9907268.
\bibitem{ref14} J. Sollfrank et al., Phys. Rev. C {\bf 55}, 392 (1997).
\end{references}
\end{document}